\documentclass[5p, authoryear]{elsarticle}
\usepackage{amsmath}
\usepackage[hidelinks]{hyperref}
\numberwithin{equation}{section}

\begin{document}
\bibliographystyle{elsarticle-harv}

\begin{frontmatter}
\title{Interpretation and approximation tools for big, dense Markov chain transition matrices in ecology and evolution}

\author[igepp]{Katja Reichel\corref{cor}}
\ead{katja.reichel@rennes.inra.fr}

\author[igepp]{Valentin Bahier}

\author[igepp]{C\'{e}dric Midoux}

\author[igepp]{Jean-Pierre Masson}

\author[igepp]{Solenn Stoeckel}

\address[igepp]{INRA, UMR1349 IGEPP, F-35653 Le Rheu, France}
\cortext[cor]{Corresponding author}

\begin{abstract}

Markov chains are a common framework for individual-based state and time discrete models in ecology and evolution. Their use, however, is largely limited to systems with a low number of states, since the transition matrices involved pose considerable challenges as their size and their density increase. Big, dense transition matrices may easily defy both the computer's memory and the scientists' ability to interpret them, due to the very high amount of information they contain; yet approximations using other types of models are not always the best solution. 

We propose a set of methods to overcome the difficulties associated with big, dense Markov chain transition matrices. Using a population genetic model as an example, we demonstrate how big matrices can be transformed into clear and easily interpretable graphs with the help of network analysis. Moreover, we describe an algorithm to save computer memory by substituting the original matrix with a sparse approximate while preserving all its mathematically important properties. In the same model example, we manage to store about 90\% less data while keeping more than 99\% of the information contained in the matrix and a closely corresponding dominant eigenvector. 

Our approach is an example how numerical limitations for the number of states in a Markov chain can be overcome. By facilitating the use of state-rich Markov chain models, they may become a valuable supplement to the diversity of models currently employed in biology. \hfill \href{run:./VisualAbstract_c.png}{\emph{Visual abstract}} $\cdot$ \href{run:./Highlights.pdf}{\emph{Highlights}}

\end{abstract}

\begin{keyword} 
discrete stochastic model \sep sparse approximation \sep eigenvector \sep network analysis \sep population genetics \sep de Finetti diagram
\end{keyword}

\end{frontmatter}

\section{Introduction} \label{Int}

Natural systems often possess inherently discrete states in space, time or both. Atoms, molecules and cells, organs, individuals, populations and taxa usually appear as distinct entities; along the time axis, the radiation cycles we use as the basis for atomic clocks, neuronal action potentials, developmental stages in an organism’s life cycle, generations and the revolutions of the earth around the sun are examples for similar patterns.

Modeling these discrete systems as such can have advantages over continuous approximations. One of the earliest examples comes from thermodynamics \citep{planck_zur_1900}, where heat emission spectra could only be predicted correctly if energy “comes in packets”, known as “quanta”. This discovery led to the new field of quantum mechanics, which provided the necessary theory for understanding the photovoltaic effect \citep{einstein_uber_1905}, thus proving essential for the invention of solar cells. In biology, the re-discovery of Mendel’s rules and thus of the “quantal” nature of genetic heritability, at about the same time as Planck's famous speech, has had a similar impact on the study of evolution as the latter's research has had on thermodynamics \citep{ewens_mathematical_2004}. While most of the objects of biological research have long been recognised as discrete (\emph{e.g.}, the word \emph{individual} literally means \emph{not dividable}, a notion very similar to that of a \emph{quantum}), we still struggle with understanding the processes, such as evolution, linking them to potential emergent properties (analogous to the physicists' heat spectra) at higher levels. Preserving the discrete nature of the natural system in our models may prove vital to scientific advance in biology.

Markov chains are a classical framework for modeling state and time discrete stochastic systems. Based on the assumption that the modeled system is \emph{memoryless} \citep[Markov property;][]{markov__1906}, the basic model equation consists in multiplying a "start" vector, providing the state of the system at a given time, with a "step" matrix. This matrix holds the transition probabilities, which depend on the model parameters and typically remain constant through time, between all possible states of the system within one time step. By analysing the transition matrix, both the "short term" transient behaviour and the "long term" limiting behaviour of the model can be studied, thus putting the matrix at the centre of attention for the biological interpretation of the results. Markov chains and other related forms of matrix-based models, such as Leslie models in population dynamics, are already widely in use \citep[e.g.\ ][]{tyvand_sexually_2007, keeling_efficient_2009, wakano_mathematical_2013}
% will put in the others
, and many textbooks detailing their mathematical properties have been written \citep[e.g.\ ][]{feller_introduction_1971, allen_introduction_2011}.

However, the use of matrix-based models is often restricted to systems with a small number of states and/or transition matrices which are sparse, i.e.\ contain many zeros \citep[compare Leslie matrices,][and many other examples]{leslie_use_1945}. This is largely due to the challenges arising from big, dense matrices: if all $n$ states are quantitatively linked among themselves, there are $n^{2}$ values to be stored and referenced in subsequent calculations, and to be accounted for in an interpretation. Thus, even with access to supercomputers, discrete models of state-rich systems can be daunting, which is why they are often either abandoned or replaced by a diffusion approximation based on the Fokker-Planck / Kolmogorov equations \citep{feller_introduction_1971,ethier_markov_1986}. As a result, the state and time discrete matrix model is turned into state and time continuous differential equations, which may impose additional limits on the parameter space, obscure relevant model properties or incur other interpretational problems (e.g.\ as discussed in \citep{gale_theoretical_1990}. The suitability of approximations merits a differentiated view, and should rather be based on the nature of the system and the desired quality of the result than on technical limitations. 

An example for a time-discrete Markov chain model with a countable finite, though potentially very large, number of discrete states is the population genetic model from \cite{stoeckel_exact_2014}. It is an extension of a classic biallelic Wright-Fisher model, based on genotype frequencies and including partial asexuality and mutation. Although a diffusion approximation is widely used for allele frequency changes in biallelic Wright-Fisher models \citep[compare][]{gale_theoretical_1990, ewens_mathematical_2004}, here this does not seem to be an equally good solution. However, since the number of states is exponentially dependent both on the population size and the number of possible genotypes in the Stoeckel-Masson model (compare equation (\ref{S-eq})), keeping the discrete framework soon leads to matrix sizes beyond the capacity of any present-day computer. As transition matrices from this model are always dense, i.e.\ contain only nonzero values, we hold that they might serve as a good example for a "worst case" in the numerical handling and interpretation of big (transition) matrices.

In this article, we suggest methods which may help in interpreting both the transient and limiting behaviour of state-rich Markov chains based on the transition matrix and its dominant eigenvector, as well as a method for approximating a dense transition matrix by a sparse substitute to facilitate regular handling on a PC. For the first part, we introduce notions from network analysis and extend them to provide clear and informative diagnostic views; for the second, we describe an algorithm which keeps a predefined percentage of information about the transient behaviour %when starting from each state 
of the system, while at the same time ensuring matrix properties which are important for the model.

\section{Model example}

The population genetic model of Stoeckel and Masson \citep{stoeckel_exact_2014} describes the evolution of genotype frequencies based on a single locus with two alleles \emph{a} and \emph{A} in a fixed-size population of diploid, partially asexual organisms. States are defined as distributions of the $N$ individuals in the population on the three possible genotypes (\emph{aa}, \emph{aA}, \emph{AA}). The transition probabilities beween the states depend on a symmetric mutation rate $\mu$ and a constant rate of asexual reproduction $c$, defined as the probability that an individual in the next generation was derived asexually from a single parent.

Transition matrices $M$ resulting from this model are generally square and dense - transitions between all states are possible in one step, although some of them (e.g. all individuals \emph{aa} to all individuals \emph{AA}) are very unlikely. The corresponding Markov chain is thus irreducible (single communicating class, no absorbing states) and aperiodic (period of all states equals one, same state possible in consecutive time steps). Since the mutation rate $\mu$ is symmetric, i.e. changes from \emph{a} to \emph{A} are just as likely as the inverse, $M$ is also partially symmetric: if the transition probabilities from one particular state to all others have been calculated, swapping the names of all alleles also gives a correct result (compare figure \ref{histos} and \ref{triA}). The notation in this article assumes left-stochastic matrices (columns represent the transition probabilities from one state to all others and thus sum to one), which implies that the limiting behaviour of the Markov chain is described by its transition matrices' (normalized) right eigenvector $v$ to the eigenvalue with the largest absolute value \cite[and multiplicity one, see][]{perron_zur_1907}, one.

\begin{table} 
\begin{tabular}{|l|l|l|l||c|c||r|} 
\hline
$N$ & $\mathcal{P}$ & $\mathcal{L}$ & $\mathcal{A}$ & $g$ & $|S|$ & memory use\\
\hline
20 & 2 & 1 & 2 & 3 & 231 & 420 KB \\
100 & 2 & 1 & 2 & 3 & 5 151 & 205 MB \\
500 & 2 & 1 & 2 & 3 & 125 751 & 120 GB \\
1000 & 2 & 1 & 2 & 3 & 501 501 & 2 TB \\
\hline
20 & 4 & 1 & 2 & 5 & 10 626 & 865 MB \\
20 & 2 & 2 & 2 & 9 & 3 108 105 & 75 TB \\
20 & 2 & 1 & 4 & 10 & 10 015 005 & 730 TB \\
20 & 2 & 2 & 4 & 100 & $9.8 \times 10^{20}$ & $6.5 \times 10^{21}$ YB\\
\hline 
\end{tabular}
\caption{Examples of matrix size based on the Stoeckel-Masson model. Memory sizes are approximate and assume 64-bit accuracy.} \label{tab1}
\end{table}

The number of states in this model, and thus the size of the transition matrix $M$, depends on the one hand on the population size and on the other hand on the complexity of the genomic system being modeled, in particular the number of different genotypes possible. For a given number of genotypes $g$, the cardinality of the state space $S$ (respective number of rows and columns in the transition matrix) 
% if I write "or", it might be understood as a logical "or" (i.e. only the number of columns OR the number of rows equals S) - hope it's ok with "respective"
in a genotype-based discrete stochastic model is:
\begin{equation}\label{S-eq} 
\left\vert{S}\right\vert = \left( \!\!\! {g \choose N}\!\!\!\right)
 = \frac
{\left( N + g-1 \right)!}
{N! \cdot \left(g-1\right)!} \\
\end{equation} 
From this equation it follows that the number of states increases exponentially with $1+ (g-1)/(N+1)$ for increasing $N$ and with $1+ N/g$ for increasing $g$. For the number of possible genotypes, the ploidy level of the organism $\mathcal{P}$, the number of (partially linked) loci $\mathcal{L}$ and their respective numbers of alleles $\mathcal{A}_{i}$, with $i \in 1 \ldots \mathcal{L}$, need to be taken into account:
\begin{equation}\label{g-eq} 
g = \prod_{i=1}^{\mathcal{L}} \left( \!\!\!{\mathcal{A}_{i} \choose \mathcal{P}}\!\!\!\right)
= \prod_{i=1}^{\mathcal{L}} 
\frac{\left(\mathcal{A}_{i} 
+ \mathcal{P} -1 \right)!}
{\mathcal{P}! \cdot
\left(\mathcal{A}_{i}-1\right)!}
\end{equation}
Examples for the size of the resulting transition matrices are given in table \ref{tab1}. From these numbers, it is clear that a realistic "base-by-base" model of a full genome is still far beyond the capacity of current computer technology; however, many cases (biallelic SNPs, unlinked loci or blocks of completely linked loci) can already be interpreted based on the very simple \emph{one-locus/two-alleles} model. It remains the dependence of $|S|$ on the population size $N$, which is fortunately not as strong (for $N>g-1$). 

To illustrate our methods, we will mostly use transition matrices derived for completely sexual populations ($c=0.0$), a case for which both transient and limiting behaviour are generally known and interpretations can be easily verified \citep{de_finetti_conservazione_1927, ewens_mathematical_2004}. For the mutation rate, $\mu = 10^{-6}$ was chosen as a plausible value based on experimental estimates \citep{kronholm_influence_2010}. $N$ is either 5 ($|S| = 21$), 20 ($|S| = 231$) or 100 ($|S| = 5 151$).

\section{Visualisation} \label{Vis}

An intuitive first step in analysing the transient behaviour of a Markov chain model is a diagnostic visualisation of the transition matrix; ideally, it can also be used later on to summarize the results in an easily accessible way, thus providing a basis for a direct biological interpretation.

\subsection{Heat map} \label{histo}

\begin{figure}
\center
\includegraphics[trim = 20mm 20mm 10mm 5mm, clip, width=0.5\textwidth]{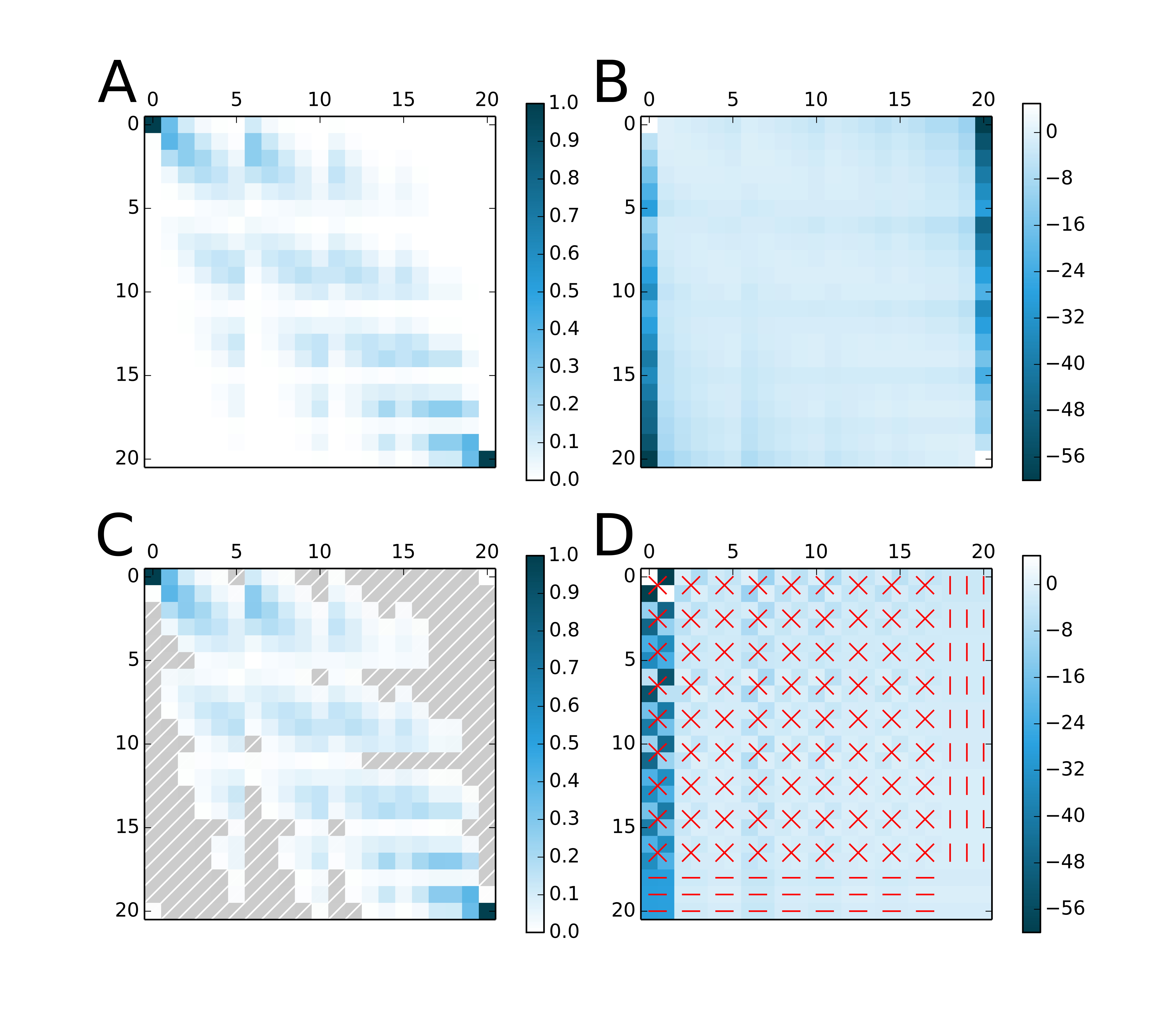}
\caption{Heat maps of transition matrices for $N=5, \mu=10^{-6}, c=0.0$. A. original probabilities, dense matrix B. logit(10) transformed probabilities, dense matrix C. sparse approximate matrix of A, implicitly stored zero values in hatched grey D. as in B, with alternative state order, red lines connect identical values. \hfill \href{run:./Fig1_histograms5_c.png}{\emph{full size}}} \label{histos}
\end{figure}

A heatmap or histogram of the transition matrix, where the transition probabilities $p$ are symbolised by colour/\allowbreak shade or height, is perhaps the easiest way to visualise it (figure \ref{histos}). In some cases, the resolution can be enhanced by an appropriate transformation of the range of values for $p$, for example by using a negative logarithm ($[0;1] \rightarrow [0; \infty]$) or a \emph{logit} transformation ($[0;1] \rightarrow [-\infty; \infty]$).

For big matrices, heat maps can be costly to produce (memory size) and are often still not very clear, due to the large number of cases. Yet they may help to recognise basic patterns (symmetries, groups of similar / more strongly connected states etc.) of potential value for finding more adapted visualisations / numerical methods.

\subsection{Network display}

The duality between matrices and graphs \citep[e.g.\ ][]{allen_introduction_2011, aghagolzadeh_transitivity_2012} opens up an alternative way for the visualisation and mathematical analysis of either structure. In a graph $\mathcal{G(V, E)}$, the states of a Markov chain are thus represented as nodes/vertices $\mathcal{V}$ and the transitions as (weighted and directed) edges $\mathcal{E}$ connecting them, which is especially useful for sparse transition matrices. 

For big, dense matrices, the amount of edges in the resulting complete multidigraph (of edge multiplicity two) 
equals the number of entries in the transition matrix and thus appears to prohibit all interpretation. We therefore developed methods, based on concepts from network theory, to selectively display edges and use the nodes to summarise information about each state of the model system. Thus a number of very clear synthetic representations can be constructed, taking into account different time scales: from one generation (based on $M$) across $t$ generations (based on $M^{t}$) up to the long-time equilibrium (dominant eigenvector of $M$, $v$).

\begin{figure}
\center
\includegraphics[trim = 20mm 20mm 10mm 5mm, clip, width=0.5\textwidth]{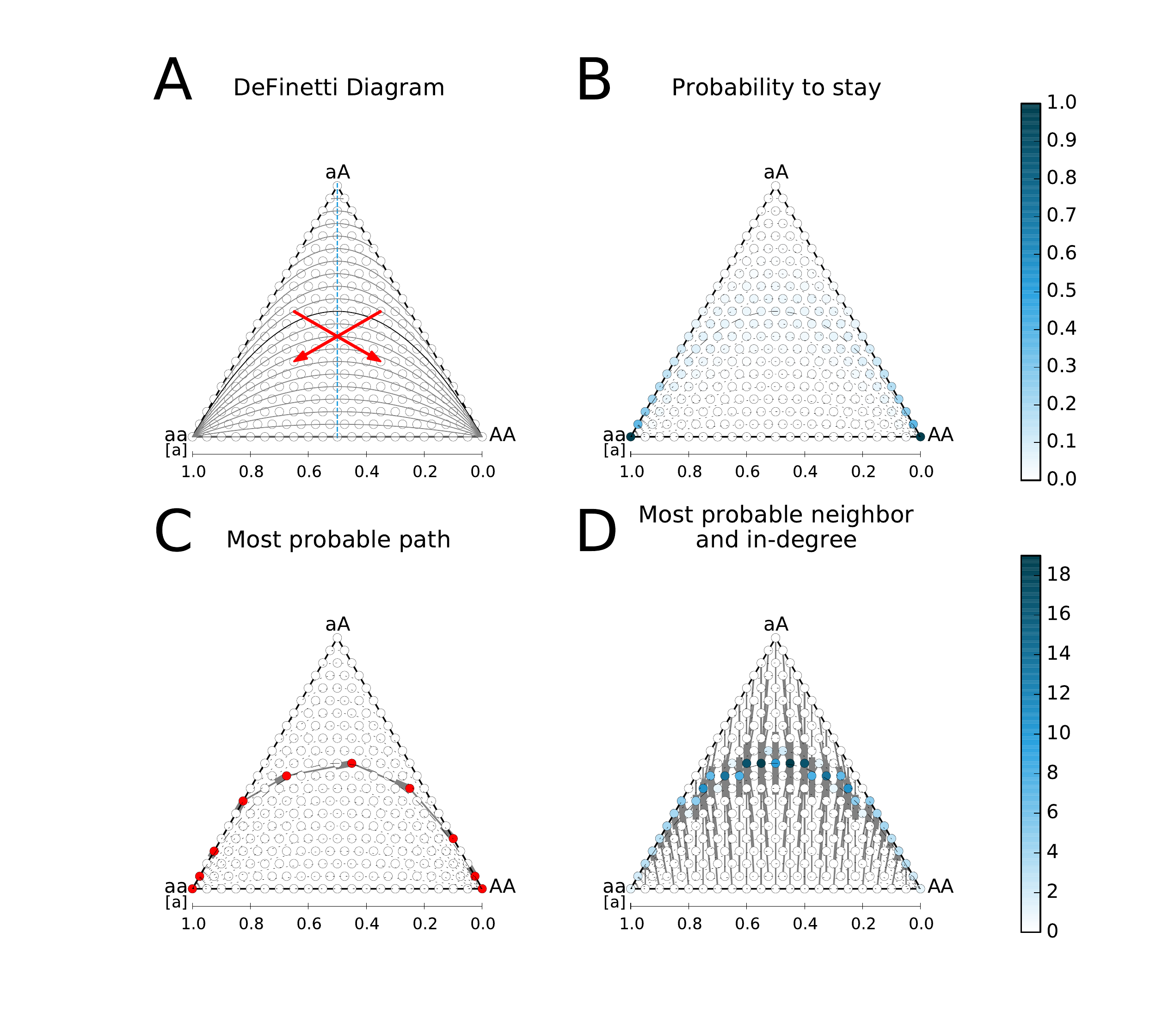}
\caption{Network display of transition matrices for $N=20, \mu=10^{-6}, c=0.0$. A. \emph{De Finetti} diagram showing symmetry (dashed blue axis, red arrows corresponding to identical probabilities) and $F_{IS}$ isocurves (gray and black) B. $p_{stay}$ (node color) C. most probable path connecting (N,0,0) to (0,0,N) D. most probable neighbors (directed edges) and in-degree (node color). \hfill \href{run:./Fig2_triangles_c.pdf}{\emph{full size}}} \label{triA}
\end{figure}

To facilitate a biological interpretation, arranging the nodes according to biological "metadata" about the states can be very important. For our example model, where states represent distributions of individuals on three genotypes (\emph{aa}, \emph{aA}, \emph{AA}) under a constant population size, we placed the nodes in a \emph{de Finetti} diagram  \citep[see figure \ref{triA}, ][]{de_finetti_conservazione_1927}, a specialised ternary plot for such population genetic data. In other circumstances, parameters such as geographic location, trophic level, functional dependence etc. may suggest "natural" orders for the states.

\subsubsection{Edges}

\paragraph{Most probable neighbor} This is the counterpart of a \emph{nearest neighbor} if distances (edge weights) represent probabilities. For each state $i$, there are one or several states $j$ which have the \emph{highest} probability to be the destination of a transition in the next time step; tracing these connections gives the expectation for the one-step transient behaviour of the model.\\
$\triangleright$ In our example, the most likely state for the next generation (figure \ref{triA}) is always on or very near to the Hardy-Weinberg Equilibrium, which is represented by the curve going through $(1/4; 1/2; 1/4)$ in the diagram.

\paragraph{Most probable path} This is the counterpart of a \emph{shortest path} if distances (edge weights) represent probabilities. For each non-commutative pair of states $i$ and $j$, there exists at least one series of consecutive edges connecting $i$ to $j$ along which the \emph{product} of the edge weights is \emph{maximal}. It can be determined by using an "ordinary" shortest path algorithm \ \citep[e.g.\ ][]{dijkstra_note_1959, biswas_generalisation_2013} on a negative \emph{log} transform of the transition matrix. The most probable path is the most likely trajectory of the model system to get from one state to another; \\
$\triangleright$ In our example (figure \ref{triA}), a change from a population with only the \emph{aa} genotype to one with only the \emph{AA} genotype would closely follow the Hardy-Weinberg curve.

\paragraph{Flow threshold} Using the smallest probability along the most likely path between two nodes $i$ and $j$ as a threshold, very rare transitions can be excluded. \\
$\triangleright$ In our example (figure \href{run:./Fig2b_triangles_c.pdf}{2b}, supplement), horizontal transitions along the base of the triangle, where no heterozygotes are produced despite of two homozygous genotypes being present in the population, would be excluded.

\subsubsection{Nodes}

\paragraph{Degree} For each node in a graph representing a dense matrix, the number of incoming (\emph{in-degree}) and outgoing (\emph{out-degree}) edges is equal to the number of nodes (matrix rows/columns). Differences only result from selective edge plotting and have to be interpreted according to context. \\
$\triangleright$ In our example (figure \ref{triA}), the nodes with the highest in-degree are nearest neighbors to the largest number of nodes; if all states were equally likely at the current generation, those next to  $(0.25; 0.5; 0.25)$ on the Hardy-Weinberg curve would be the most likely in the next generation.

\paragraph{Betweenness-centrality} Based on the same concept as the \emph{most probable path}, this can be redefined as the number of \emph{most probable paths} passing through each node when connections between each pair of nodes are considered. It can be derived in a similar way as the \emph{most probable path}, by applying a standard algorithm developed for additive distances to a negative $log$ transform of the multiplicative probabilities in $M$. Nodes with a high betweenness-centrality represent frequent transient states.\\
$\triangleright$ In our example, these are all the states along the Hardy-Weinberg curve except for the fixation states (figure \href{run:./Fig2c_triangles_c.pdf}{2c}, supplement).

\paragraph{Probabilities} For each state $i$ in the Markov chain model, several probabilities can be calculated - and displayed on the nodes - to describe both the transient and limiting behaviour:

\begin{itemize}
\item[$p_{stay}$] – \emph{probability to stay for one time step}\\
$p_{stay}(i) = p_{i,i}$, the probabilities on the matrix diagonal; for each state $i$ this is the probability that the system remains at state $i$ for the next time step ("stickiness"). This probability allows the easy detection of (near-)absorptive states. \\
$\triangleright$ In population genetics, the fixation states $\lbrace(N;0;0),$  $(0;0;N)\rbrace$ are typical examples (figure \ref{triA}).

\item[$p_{out}$] – \emph{probability to leave in one time step} \\
$p_{out}(i) = 1-p_{i,i}$, the column sums of the matrix without the diagonal; for each state $i$ this is the probability that the system changes state at the next time step ("conductivity"). Being the opposite of $p_{stay}$, this probability allows the detection of states which are rarely occupied for consecutive time steps.\\
$\triangleright$ In our example, these are the states where the population consists of an approximately even mixture of both homozygotes (central basis of the triangle) or only of heterozygotes (top of the triangle; figure \href{run:./Fig2b_triangles_c.pdf}{2b}, supplement).

In contrast, the row sums of a left-stochastic matrix may exceed one and are thus not probabilities. As a result of the Markov property, a \emph{probability to arrive} always depends on the state at the previous time step, which results in a number of possible definitions.

\item[$p(i|j)$] \emph{probability to arrive from state $j$ in one time step} \\
$p(i|j) = p_{j, i}, j \in S$, all probabilities in one column of the transition matrix; the probability distribution (mean, variance, skew according to arrangement of nodes) for transitions starting from one particular state. This allows the prediction of the most likely states for the next time step. \\
$\triangleright$ In our example, the variance around the fixation states is much more limited than at the interior states of the triangle (figure \href{run:./Fig2c_triangles_c.pdf}{2c}, supplement).

\item[$p_{in}$] \emph{probability to arrive in one time step} \\
$p_{in}(i) = 1/(|S|-1) \cdot \sum_j p_{j, i}$ for $i \neq j$, the row sums of the matrix divided by the number of states; probabilities to arrive at state $i$ if all previous states are equally likely. This shows states which are generally very likely destinations for one-step transitions.\\
$\triangleright$ In our example, these are the states around the Hardy-Weinberg curve (figure \href{run:./Fig2b_triangles_c.pdf}{2b}, supplement).
 
\item[$p_{in}^{\infty}$] \emph{probability to arrive in an infinite run} \\
$p_{in}^{\infty}(i) = \sum_j p_{j, i} \cdot v_{j} $ for $i \neq j$, the sum over the element-wise product of eigenvector and matrix row, without the diagonal; probabilities to arrive at state $i$ if the likelihood of the previous states is distributed according to the limiting distribution. This shows the states which are the most frequent destination of transitions in an infinite run of the model.\\
$\triangleright$ In our example, these are the two states next to the fixation states where there is exactly one "foreign" allele (figure \href{run:./Fig2c_triangles_c.pdf}{2c}, supplement).

\item[$p^{\infty}$] \emph{limiting distribution / eigenvector-centrality} \\
$p^{\infty}(i) = v_{i}$, the eigenvector; probability to find the system at state $i$ after infinitely many time steps, or proportion of time spent in each state averaged over infinitely many time steps (limiting distribution). This is the prediction for the most likely states independently of the start state.\\
$\triangleright$ As is well known for our example, these are the fixation states (figure \href{run:./Fig2b_triangles_c.pdf}{2b}, supplement).

\end{itemize}

\paragraph{Others} For each modeled system, there might also be indices which are more specific to the scientific questions behind it. In our example, one such index is the
expected time to fixation $E(t_{fix})$, which can be easily derived if the fixation states are considered absorptive \citep{allen_introduction_2011}. \\
$\triangleright$ For our example, the resulting graph in figure \href{run:./Fig2c_triangles_c.pdf}{2c} shows that the expected time to fixation depends predominantly on the current state's allele frequencies.

\section{Approximation} \label{Apx}

While the visualisation methods described in the previous section may help to structure and interpret data, they do not solve the memory size problem. On the contrary, some methods which involve eigenvector calculation ($p_{in}^{\infty}$, $p^{\infty}$) or finding the inverse of a matrix ($E(t_{fix})$) are computationally expensive and may need a lot of time; this is further exacerbated by the limited availability of RAM as it is shared between the matrix and the algorithm's intermediate results. As we have seen (table \ref{tab1}), some matrices also exceed the size of the RAM, but may yet be stored on a hard drive instead (serialisation); here calculations are even slower, since there are increased access times on top.

Multiple ways exist for increasing the maximum possible number of states in the model while keeping a better balance between speed and matrix size. One example consists in "virtualising" the matrix by iteratively calculating only those parts needed for a particular task (e.g.\ multiplication with a vector) without ever storing the entirety of all entries simultaneously. Whether this approach is faster than hard-drive storage depends on the hardware used and on the numerical complexity of constructing the matrix. Speed gains may be achievable by parallelisation, or even by simply reordering the states according to matrix symmetries (figure \ref{histos}). 

The method we present here takes a different route: limiting the amount of values to be stored by substituting a dense matrix with a sparse approximate having generally the same mathematical properties. Most near-zero values in the matrix - except for some which assure aperiodicity and irreducibility - will be rounded to zero and the remaining values rescaled to obtain a left-stochastic matrix again, thus making it possible to save memory space by omitting the zero entries. In contrast to, e.g., an approximation based on a flow threshold, our method could also be combined with iterative matrix calculation to construct the sparse approximate matrix directly or perform mathematical operations with a "virtual" sparse matrix.

The algorithm iterates over all columns of the transition matrix $M$ and excludes (almost) all values which, in total, contribute less than a threshold value $s \in [0,1]$ to the column sum:
\begin{itemize}
\item for all columns $C^i = M_{1\ldots |S|,i}$ with $i \in [0, |S|]$:
\begin{enumerate}
\item create a permutation $R$ of the row indices so that the corresponding entries are ranked according to size:\\
$R \leftarrow$ \emph{ordinalrank}$(j\, |\, 1 \geq C^i_{j} \geq 0)$

\item find the minimal rank (index of $R$) so the corresponding entries sum at least to the threshold value $s$ \\
$r \leftarrow min(k)$ for $\sum_{R_1}^{R_k} C^i_{R_k} \geq s $

\item keep at least the two biggest values per column\\
$r \leftarrow max(2,r)$

\item keep all values of equal rank \\
while $C^i_{R_{r+1}} = C^i_{R_{r}}$ : $r \leftarrow r+1$

\item round all values with ranks greater then $r$ to zero, but keep those on the main diagonal and its neighbors \\
$C^i_{R_k} \leftarrow 0$ for all $k$ with \\
$k > r \vee R_k \notin \{(i-1, i, i+1) \, mod \, |S|\}$

\item rescale the column to sum to $1$ \\
$C^i \leftarrow C^i/sum(C^i)$.

\end{enumerate}

\end{itemize}

\begin{figure}
\center
\includegraphics[trim = 0mm 20mm 0mm 0mm, clip, width=0.5\textwidth]{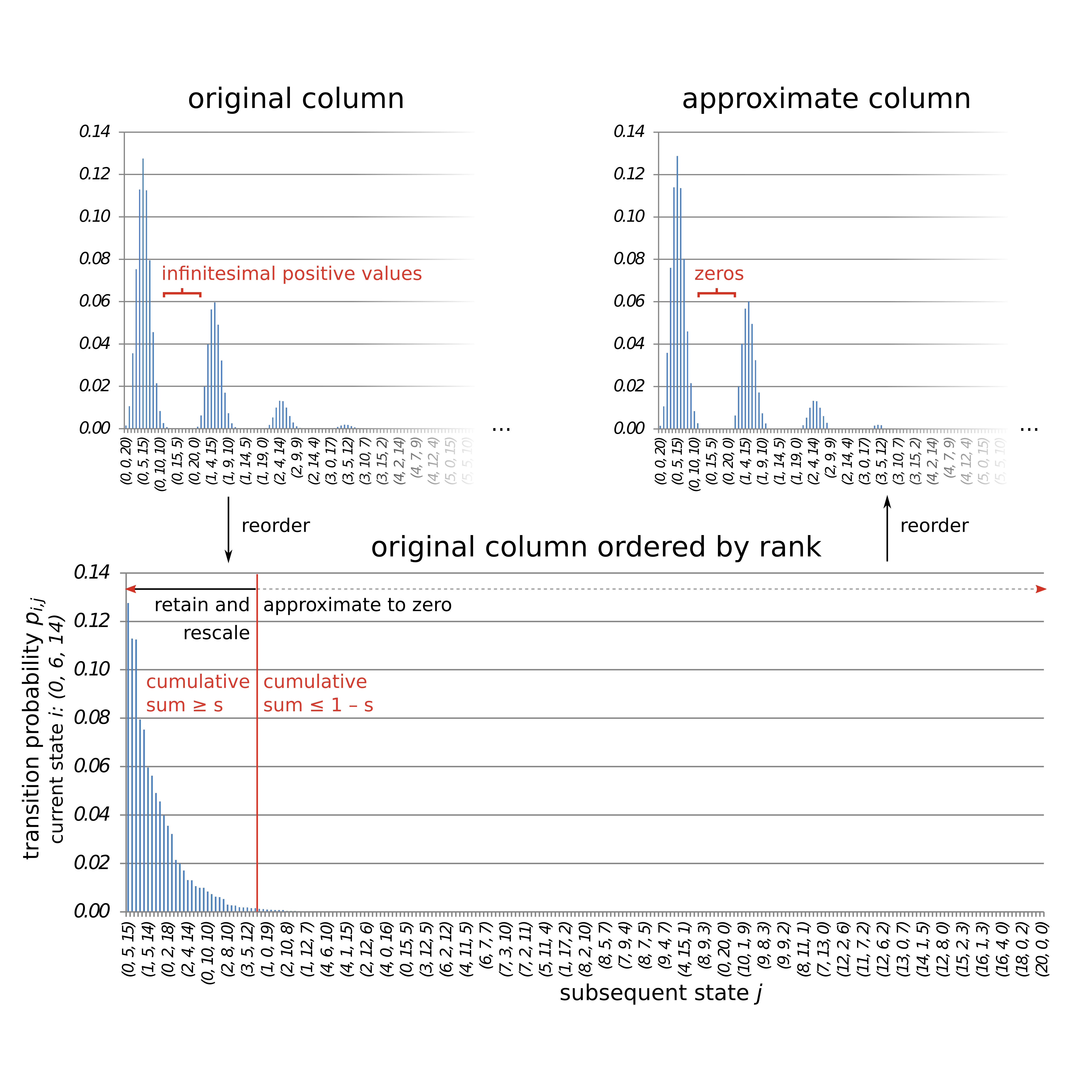}
\caption{Illustration of the approximation algorithm $(s=0.99)$ for $N=20, \mu=10^{-6}, c=0.0$ and the state $(0,6,14)$. Reordering is based on the relative size of the column entries and their index in the original column, respectively. \hfill \href{run:./Fig3_ApproximationAlgorithm_c.png}{\emph{full size}}} \label{Algo}
\end{figure}

The first two steps, together with the rounding in step five, form the core of the algorithm (compare figure \ref{Algo}), steps three and four prevent distortions and the others ensure the continued validity of those matrix properties we considered essential in the context of Markov chains. Irreducibility is assured by keeping at least one outgoing and one incoming transition probability (step five), aperiodicity by keeping the main diagonal (step five), and the rescaling of each column ensures left-stochasticity of the matrix (step six). On the contrary, the property that one-step transitions are possible between all states is deliberately given up. 

Both the efficiency (density of the resulting matrix) and the bias vary according to the value of $s$ and the distribution of values in the original matrix. If $s$ is low or the probability distribution in the column is highly uneven, more values will be discarded (compare figure \ref{Algo}); since $s$ has to be determined heuristically, we recommend testing successively increasing values.

Different ways of estimating the bias introduced by this approximation method are possible. The sum of the difference between the entries of the approximate and original matrices has a theoretical upper limit of $(1-s) \cdot |S|$. Alternatively, we used the bias in the limiting distributions resulting from approximate and original matrix as a criterion: a population genetic parameter commonly cited as a reference in estimating the rate of asexual reproduction is $F_{IS}$ \citep[e.g.\ ][]{halkett_tackling_2005}, thus we were interested in determining the effect of the approximation on the long-term probability distribution of $F_{IS}$. 

\begin{figure}
\center
\includegraphics[trim = 20mm 10mm 10mm 5mm, clip, width=0.5\textwidth]{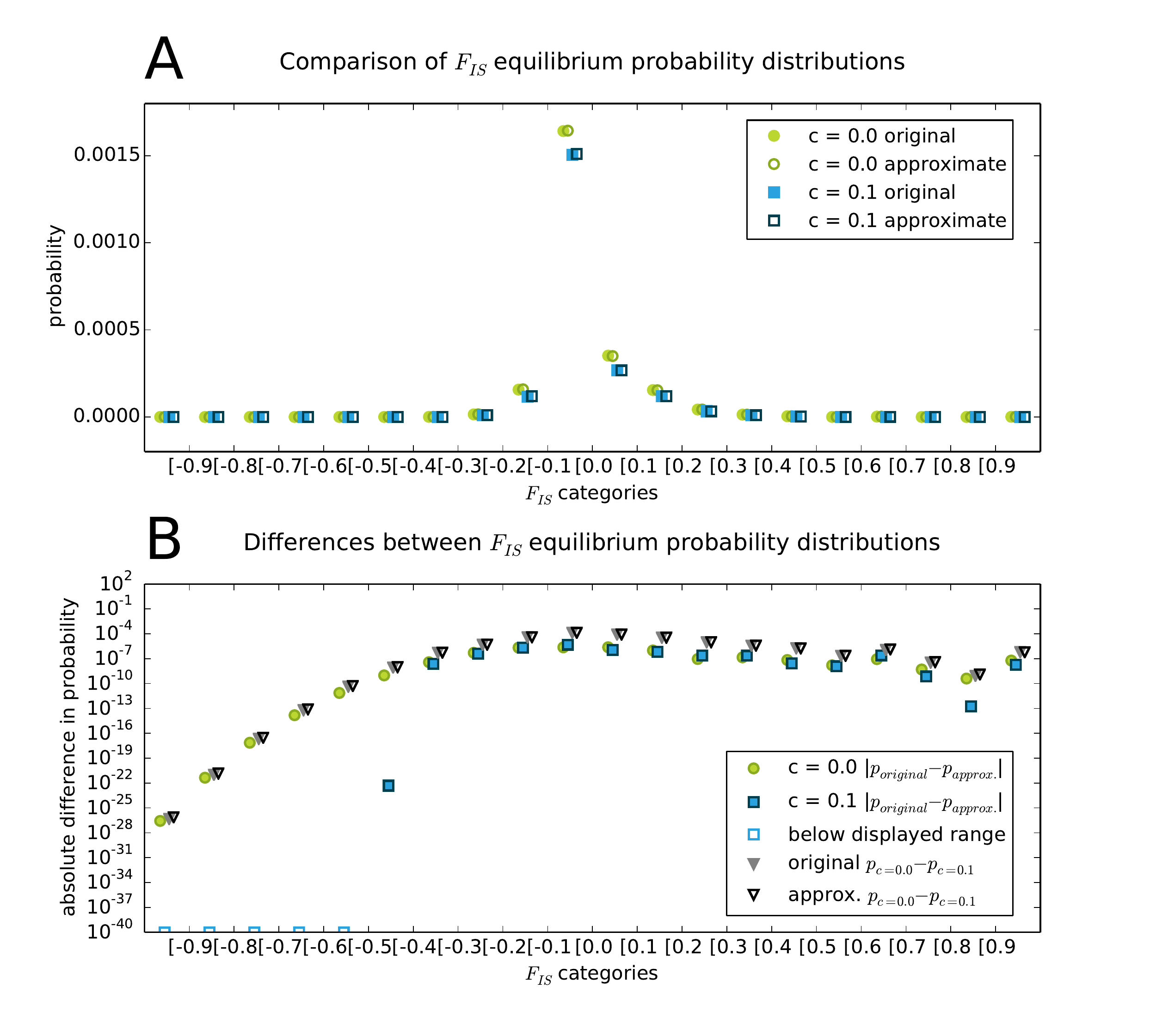}
\caption{Comparison of the limiting distribution of $F_{IS}$ for $N=100, \mu=10^{-6}, c=\{0.0, 0.1\}$. A. probability distributions based on the original (filled symbols) and the approximate (unfilled symbols) matrix  B. pairwise differences between probability distributions, biologically interesting distances marked by triangles. \hfill \href{Fig4_Fiscomp100_c.pdf}{\emph{full~size}} } \label{Fisc}
\end{figure}

The value of $s$ can be optimised so that the bias of the approximation does not interfere with the biological interpretation of results. As can be seen in figure \ref{Fisc}, the differences in the long-term expected distribution of $F_{IS}$ between two values of $c$, determined from the approximate matrices, closely follow those obtained from the original matrices. The example is based on a case (population size big, mutation rate small and asexuality rare) where the expected differences are extremely small – the \emph{p-value} derived from a \emph{G-test} of the two distributions is at the order of $10^{-6}$ (original matrices) to $10^{-3}$ (approximate matrices) - but even after rounding more than 92\% ($s = 0.99$) of the matrix entries to zero, the $F_{IS}$ distributions remain largely unchanged. While the approximate matrices are less suitable for the fine-scale quantitative analysis of rare cases (e.g. left and right tails of the distribution in figure \ref{Fisc}), they still provide sufficiently accurate probability distributions to allow a correct biological interpretation.

\section{Discussion}

While computational models involving big, dense matrices still remain a challenge, the difficulties are not necessarily insurmountable. As we have shown, some basic tools such as network theory and sparse data formats may be sufficient to allow the calculation, visualisation and interpretation of dense, state-rich Markov chain transition matrices. Our commented source code is freely available and may easily be adapted to fit the requirements of other models. Combinations with other approaches, e.g. parallelisation or the use of algorithms for sparse matrices \citep[inspired by][or others]{gambin_aggregation_2001, busic_bounded_2012}, are also possible. This opens up new horizons for the description of state-rich discrete stochastic models in ecology and evolutionary biology, providing an alternative to diffusion approximations for situations when these are not suitable.

Using the conceptual likeness between Markov chains and networks appears to be a promising route towards an effective tool in interpreting state-rich models. The representations we found provide results which are congruent to those obtained from previous models \citep{de_finetti_conservazione_1927, ewens_mathematical_2004}, with the additional benefit of providing a sense for the expected natural variation due to the stochasticity of the model. While \emph{de Finetti} diagrams are rather specific to our example, representing Markov Chains by networks is not and many other layouts are possible. Efficient illustrations are not a substitute for strict mathematical analysis, yet can be a guide and reference in the process.

Sparse approximations of big, dense transition matrices may be an additional way to overcome technological limitations. However, both effectiveness and bias of the approximation are largely dependent on the matrix entries. Using our algorithm, the approximation accuracy can be sufficiently increased by changing the parameter $s$, while at the same time allowing a very high efficiency due to a pronounced skew in the probabilities within each column of the sample matrix. To estimate the bias, especially in derived values such as $F_{IS}$, it is still necessary to calculate the original matrix once for the sake of comparison. Otherwise, the approximate matrix can be directly constructed by iterating over columns.

Individual-based models are becoming more and more popular in biology \citep{black_stochastic_2012}, which will further increase the frequency of encountering computationally challenging cases such as the one we used as our example. In population genetics, modeling more complex evolutionary parameters such as life cycles and reproductive mechanisms, multi-dimensional fitness landscapes or dispersal may often lead to the necessity of extending the traditional models from allele frequencies \citep{ewens_mathematical_2004} to genotypes. Due to the diploid/polyploid nature of most higher organisms, this will necessarily increase the size of transition matrices and equation systems to be analysed. By presenting our approach, we hope to encourage and inspire others to extend and adapt our methods, thus further paving the way for the use of Markov Chain models with big, dense transition matrices.

\section*{Acknowledgements} 

We thank J\"{u}rgen Angst, Sophie Arnaud-Haond, Florent Malrieu, Nicolas Parisey and Fran\c{c}ois Timon for constructive discussions, and all reviewers for their helpful criticism. This study is part of the CLONIX project (ANR-11-BSV7-007) financed by the French National Research Agency. Katja Reichel receives a PhD grant by the R\'{e}gion Bretagne and the division “Plant Health and Environment” of the French National Institute of Agricultural Research (INRA).

\section*{Data accessibility}

We implemented all algorithms in Python 2.7 and 3.4, using in particular the extension modules numpy/scipy, matplotlib and networkx (\cite{oliphant_python_2007}, \cite{hunter_matplotlib:_2007}, \cite{hagberg_exploring_2008}). Our code, including documentation, is collected in the module \href{run:./mamoth.zip}{mamoth} (supplement), also available from: \url{http://www6.rennes.inra.fr/igepp\_eng/Productions/Software} .

\section*{References}

\bibliography{MMP} 

\begin{thebibliography}{26}
\expandafter\ifx\csname natexlab\endcsname\relax\def\natexlab#1{#1}\fi
\expandafter\ifx\csname url\endcsname\relax
  \def\url#1{\texttt{#1}}\fi
\expandafter\ifx\csname urlprefix\endcsname\relax\def\urlprefix{URL }\fi

\bibitem[{Aghagolzadeh et~al.(2012)Aghagolzadeh, Barjasteh, and
  Radha}]{aghagolzadeh_transitivity_2012}
Aghagolzadeh, M., Barjasteh, I., Radha, H., 2012. Transitivity matrix of social
  network graphs. In: Statistical Signal Processing Workshop ({SSP)}, 2012
  {IEEE}. {IEEE}, pp. 145--148.

\bibitem[{Allen(2011)}]{allen_introduction_2011}
Allen, L. J.~S., 2011. An introduction to stochastic processes with
  applications to biology, 2nd Edition. Chapman \& Hall, Boca Raton ({FL)}.

\bibitem[{Biswas et~al.(2013)Biswas, Alam, and
  Doja}]{biswas_generalisation_2013}
Biswas, S.~S., Alam, B., Doja, M.~N., 2013. Generalisation of {D}ijkstra's
  algorithm for extraction of shortest paths in directed multigraphs. Journal
  of Computer Science 9~(3), 377--382.

\bibitem[{Black and McKane(2012)}]{black_stochastic_2012}
Black, A.~J., McKane, A.~J., 2012. Stochastic formulation of ecological models
  and their applications. Trends in Ecology \& Evolution 27~(6), 337--345.

\bibitem[{Bu\v{s}ic et~al.(2012)Bu\v{s}ic, Djafri, and
  Fourneau}]{busic_bounded_2012}
Bu\v{s}ic, A., Djafri, H., Fourneau, J., 2012. Bounded state space truncation
  and censored {M}arkov chains. In: 2012 {IEEE} 51st Annual Conference on
  Decision and Control ({CDC)}. pp. 5828--5833.

\bibitem[{de~Finetti(1927)}]{de_finetti_conservazione_1927}
de~Finetti, B., 1927. Conservazione e diffusione dei caratteri mendeliani. nota
  i. caso panmittico. In: Rendiconti della R. Accademia Nazionale dei Lincei.
  Vol. V (11-12). pp. 913--921.

\bibitem[{Dijkstra(1959)}]{dijkstra_note_1959}
Dijkstra, E.~W., 1959. A note on two problems in connexion with graphs.
  Numerische Mathematik 1, 269--271.

\bibitem[{Einstein(1905)}]{einstein_uber_1905}
Einstein, A., 1905. {\"U}ber einen die {E}rzeugung und {V}erwandlung des
  {L}ichtes betreffenden heuristischen {G}esichtspunkt. Annalen der Physik
  322~(6), 132--148.

\bibitem[{Ethier and Kurtz(1986)}]{ethier_markov_1986}
Ethier, S.~N., Kurtz, T.~G., 1986. Markov processes: characterization and
  convergence. Wiley.

\bibitem[{Ewens(2004)}]{ewens_mathematical_2004}
Ewens, W.~J., 2004. Mathematical Population Genetics: I. Theoretical
  Introduction, 2nd Edition. Interdisciplinary applied mathematics. Springer,
  New York.

\bibitem[{Feller(1971)}]{feller_introduction_1971}
Feller, W., 1971. An introduction to probability theory and its applications.
  Wiley.

\bibitem[{Gale(1990)}]{gale_theoretical_1990}
Gale, J.~S., 1990. Theoretical Population Genetics. Springer.

\bibitem[{Gambin and Pokarowski(2001)}]{gambin_aggregation_2001}
Gambin, A., Pokarowski, P., 2001. Aggregation algorithms for {M}arkov chains
  with large state space. In: Computer Algebra in Scientific Computing {CASC}
  2001 - Proceedings of the Fourth International Workshop. Springer, pp.
  195--212.

\bibitem[{Hagberg et~al.(2008)Hagberg, Schult, and
  Swart}]{hagberg_exploring_2008}
Hagberg, A.~A., Schult, D.~A., Swart, P.~J., 2008. Exploring network structure,
  dynamics, and function using {NetworkX}. In: Proceedings of the 7th Python in
  Science Conference ({SciPy2008)}. Pasadena ({CA)}, {USA}, pp. 11--15.

\bibitem[{Halkett et~al.(2005)Halkett, Simon, and
  Balloux}]{halkett_tackling_2005}
Halkett, F., Simon, J., Balloux, F., 2005. Tackling the population genetics of
  clonal and partially clonal organisms. Trends in Ecology \& Evolution 20~(4),
  194--201.

\bibitem[{Hunter(2007)}]{hunter_matplotlib:_2007}
Hunter, J.~D., 2007. Matplotlib: a {2D} graphics environment. Computing in
  Science \& Engineering 9~(3), 90--95.

\bibitem[{Keeling and Ross(2009)}]{keeling_efficient_2009}
Keeling, M.~J., Ross, J.~V., 2009. Efficient methods for studying stochastic
  disease and population dynamics. Theoretical Population Biology 75~(2--3),
  133--141.

\bibitem[{Kronholm et~al.(2010)Kronholm, Loudet, and
  Meaux}]{kronholm_influence_2010}
Kronholm, I., Loudet, O., Meaux, J.~d., 2010. Influence of mutation rate on
  estimators of genetic differentiation - lessons from \emph{Arabidopsis
  thaliana}. {BMC} Genetics 11~(1), 33.

\bibitem[{Leslie(1945)}]{leslie_use_1945}
Leslie, P.~H., 1945. On the use of matrices in certain population mathematics.
  Biometrika 33~(3), 183--212.

\bibitem[{Markov(1906)}]{markov__1906}
Markov, A.~A., 1906. Extension of the limit theorems of probability theory to a
  sum of variables connected in a chain. Proceedings of the Society of Physics
  and Mathematics at the University of Kazan 15~(2), 135--156.

\bibitem[{Oliphant(2007)}]{oliphant_python_2007}
Oliphant, T.~E., 2007. Python for scientific computing. Computing in Science \&
  Engineering 9~(3), 10--20.

\bibitem[{Perron(1907)}]{perron_zur_1907}
Perron, O., 1907. Zur {T}heorie der {M}atrices. Mathematische Annalen 64~(2),
  248--263.

\bibitem[{Planck(1900)}]{planck_zur_1900}
Planck, M., 1900. Zur {T}heorie des {G}esetzes der {E}nergieverteilung im
  {N}ormalspectrum. In: Verhandlungen der Deutschen Physikalischen
  Gesellschaft. Vol.~2. pp. 237--245.

\bibitem[{Stoeckel and Masson(2014)}]{stoeckel_exact_2014}
Stoeckel, S., Masson, J.-P., 2014. The exact distributions of {FIS} under
  partial asexuality in small finite populations with mutation. {PLoS} {ONE}
  9~(1), e85228.

\bibitem[{Tyvand and Thorvaldsen(2007)}]{tyvand_sexually_2007}
Tyvand, P.~A., Thorvaldsen, S., 2007. A sexually neutral discrete {M}arkov
  model for given sum males + females. Theoretical Population Biology 72~(1),
  148--152.

\bibitem[{Wakano et~al.(2013)Wakano, Ohtsuki, and
  Kobayashi}]{wakano_mathematical_2013}
Wakano, J.~Y., Ohtsuki, H., Kobayashi, Y., 2013. A mathematical description of
  the inclusive fitness theory. Theoretical Population Biology 84, 46--55.

\end{thebibliography}

\end{document}